# Symmetry- and gradient-enhanced Gaussian process regression for the active learning of potential energy surfaces in porous materials

View Online   Export Citation   CrossMark

Johannes K. Krondorfer, 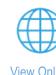 Christian W. Binder,  and Andreas W. Hauser[a]

**AFFILIATIONS**
Institute of Experimental Physics, Graz University of Technology, Petersgasse 16, 8010 Graz, Austria

**Note:** This paper is part of the JCP Special Topic on Machine Learning Hits Molecular Simulations.
[a]Author to whom correspondence should be addressed: andreas.w.hauser@gmail.com

**ABSTRACT**
The theoretical investigation of gas adsorption, storage, separation, diffusion, and related transport processes in porous materials relies on a detailed knowledge of the potential energy surface of molecules in a stationary environment. In this article, a new algorithm is presented, specifically developed for gas transport phenomena, which allows for a highly cost-effective determination of molecular potential energy surfaces. It is based on a symmetry-enhanced version of Gaussian process regression with embedded gradient information and employs an active learning strategy to keep the number of single point evaluations as low as possible. The performance of the algorithm is tested for a selection of gas sieving scenarios on porous, N-functionalized graphene and for the intermolecular interaction of $CH_4$ and $N_2$.

## I. INTRODUCTION

The ability to predict the forces and energies of molecules in the proximity of weakly interacting external structures, including surfaces, membranes, or porous materials such as metal organic frameworks and zeolites, is important for the description of a wide variety of industrially relevant processes.[1] Examples include hydrogen storage and purification,[2–4] $CO_2$ capture and sequestration,[5,6] molecular sieving technologies,[7–10] and the chiral resolution of racemic mixtures for pharmaceutical applications.[11–14] Typical approaches to calculate the macroscopic quantities of interest, such as permeation rates, selectivities, diffusion constants, adsorption energies, or gas storage capacities, are Monte Carlo-based algorithms,[15–19] molecular dynamics simulations,[20–24] or direct partition sum integration methods[25,26]—all of which are making extensive use of potential energy surface (PES) calculators at varying levels of theory.

Often, the necessary single point evaluations are performed using computationally less expensive empirical potentials and force fields, which have an analytically closed form, provide gradient information at minimal cost, and are available for any type of molecular environment. However, reduced computational costs come at the risk of insufficient accuracy or even a complete inability to capture the complicated topography of the actual PES.[13] The lack of an electronic structure treatment is the obvious reason for systematic failures in these cost-effective approaches. However, methods such as density functional theory (DFT) or wavefunction-based approaches are computationally very demanding, which renders the uninformed, point-wise exploration of the underlying PES often an unfeasible mission.

Much work has, therefore, been devoted to the development of efficient approximation schemes exploiting machine learning (ML) techniques to reduce the computational effort of PES evaluations. In general, three fundamentally different approaches can be distinguished. The first group of methods is concerned with a complete methodological replacement of computationally expensive PES evaluations by the direct and complete knowledge of the relevant energy-geometry relationship. Obviously, representatives of this branch must employ highly flexible learning techniques such as deep neural networks and necessitate considerable amounts of high-quality data, typically obtained via DFT, for training purposes.[27–31] In the second group, one attempts to correct the PES constructed at a lower level of theory in order to obtain energies with the accuracy of expensive electronic structure methods.[32–36] The third group consists of methods for an informed PES exploration given a set of data points already evaluated at a high level of theory, but with the aim to keep the number of additional evaluations as low as possible. Typically, datasets grow dynamically by single-point (SP) evaluations that are selected "on-the-fly," i.e., on demand, during an ongoing surface exploration. Motivations for these implementations are accelerated geometry optimizations,[37–39] efficient transition state searches,[40–46] or the improvement of local PES scans for better prediction of molecular properties, such as the accuracy of reaction rates within instanton rate theory.[47–49]

This article is concerned with a method of the third kind, specifically designed for molecular problems, where a separation of the molecular system into a single molecule or "mobile phase" and its preferably highly symmetrical structural environment is applicable. In contrast to generic methods focusing on inter-molecular[50,51] or intra-molecular interactions,[52] our method is trimmed for maximum efficiency in cases where the physically interesting subspace of the PES is spanned by the translational degrees of freedom of a rigid molecular object and its orientation within a porous, highly symmetric environment. The algorithm makes its own choice on where to place its SP evaluations based on three ingredients: the current set of SP evaluations on the PES of interest, the symmetry of the total molecular system, and energy gradients, since many PES calculators are providing them by default. It uses Gaussian Process Regression (GPR) to explore the PES of a molecule in a rigid rotor approximation and a geometrically "frozen" external structure. In this specific situation, it is possible to make extensive use of molecular symmetries by employing a symmetry-adapted kernel. The advantage of symmetry-adaptation has already been demonstrated for kernel ridge regression models[52–54] and PES fitting, in general.[55] In addition, the incorporation of gradient information has proven useful in the past.[28,52] GPR models for molecular PES fitting were successfully employed, and the possibility of introducing, e.g., permutation symmetries in GPR models has been demonstrated.[50,56] It is also highly suitable for the task of active learning.[51] Different point search algorithms have been proposed, ranging from simple variance optimization[57] to maximum information gain.[58] However, contrary to common GPR formulations for molecular systems, the approach presented here works without hyper-parameter scaling, a challenging necessity for algorithms formulated in internal coordinates.[39] Symmetry-adaptation, active learning, and the inclusion of gradient information are combined to develop an efficient regression model specifically suitable for molecular problems where a separation of the molecular system into a single molecule or "mobile phase" and its preferably highly symmetrical structural environment is applicable.

This article is structured as follows: First, we give a detailed overview of our GPR ansatz and its extension with respect to symmetry and gradient information. We discuss hyperparameter optimization, introduce an active learning strategy, and present the algorithm in compact form. In Sec. III, the method is then tested on four molecular systems and compared to standard GPR. Three benchmark scenarios are related to molecular sieving via a single sheet of N-substituted porous graphene. In the last test, we attempted a cost-effective fit of the high-dimensional PES describing the inter-molecular interaction of $CH_4$ and $N_2$, similar to systems discussed in Ref. 50.

## II. METHODS

The PES of a molecule in an external structure can be of very complicated topography. Yet, the obvious and meaningful discrimination between a "mobile phase" and its structural background justifies a separated view on the possible degrees of freedom and leads to a separation into inter- and intra-nuclear contributions, with the latter belonging to either the molecule or its structural environment. In this article, we assume the molecular geometry of the mobile phase, as well as the external, surrounding structure, to be fully rigid. As will be shown, this allows for tremendous exploitation of the symmetries of both subsystems, which can be used to reduce the number of single point energy evaluations in the course of a PES exploration by orders of magnitude. Our ansatz is particularly powerful in cases where the surrounding environment, e.g., a nanopore or nanocavity, is highly symmetric. Perfect applications are problems of molecular sieving or gas transport in nanoporous materials. We admit that rigidity is a strong constraint and might reduce the applicability of our approach to minimally interacting systems at first sight. However, there are three important arguments in our favor:

1. First, many intramolecular degrees of freedom do not reduce the actual molecular symmetry; for example, two out of three internal vibrations of water keep its $C_{2v}$ symmetry intact. This means that the same efficient evaluation strategy can be exploited for these modes. Even if the symmetry is reduced, e.g., by the asymmetric stretch in the case of $H_2O$, the remaining symmetry (in this case $C_s$) can still be exploited.
2. Second, the intramolecular degrees of freedom are usually of limited amplitude (unless there is a chemical reaction taking place), and a PES exploration in these modes can be added with little effort since these points are "close" in phase space, i.e., not far from the symmetric minimum geometry.
3. Third, even if intramolecular degrees of freedom may break or reduce the symmetry of the fragment, at least a large, thermodynamically relevant subspace of the total PES space is explored with maximum efficiency. Symmetry-reducing degrees can be added later as required.

Even within the "rigid rotor" approximation, the remaining six-dimensional configuration space of a molecule consisting of $N_m$ atoms in a fully rigid environment still forms a complicated sub-manifold of the $3N_m$-dimensional Cartesian space. As a first step, we will introduce a proper parametrization of this rigid rotor sub-manifold and discuss its symmetry properties. We then present a mapping of the PES that emphasizes regions of interest, i.e., thermodynamically accessible areas of lower energy. In the last step, we apply our Gaussian process regression ansatz to this transformed PES.

### A. Choice of coordinates and symmetry considerations

A rigid rotor molecule, built from $N_m$ atoms and embedded in an external rigid structure, can be described by six coordinates: its center-of-mass position and three Euler angles. We refer to this coordinate system as the rigid rotor (RR) coordinate system. Obviously, variations in these six coordinates will only explore a sub-manifold of the actual, total PES but a sub-manifold that exhibits highly useful symmetry properties. Note that the actual choice of the coordinate system is of minor relevance in this intermediate step as it is only needed to generate sample points lying exclusively on the sub-manifold. Later on, for the actual regression, these points will be expressed in Cartesian coordinates again.

The external structure might be a molecule itself, but it can also be a surface, a porous membrane, or a three-dimensional cavity. The molecule and the external structure might possess point-group or even space-group symmetries. Within their own respective Cartesian sub-spaces, any symmetry operation of either molecule or environment can be represented via the general isometric affine linear form $T = (M, b)$, with $M$ denoting an orthogonal matrix and $b$ a shift vector. The total set of valid symmetry operations is, thus, given by the composition of molecular and environmental symmetry transformations.

The group properties and the isometry of those symmetry operations have beneficial implications for calculating the distance between two points in the RR sub-manifold: For two different positions $x$ and $x'$ of a rigid molecule with respect to its environment, expressed as $3N_m$-dimensional Cartesian vectors, and with operations $T$ and $T'$ as the elements of the symmetry group, one finds

$$\|Tx - T'x'\| = \|x - T^{-1}T'x'\| = \|x - T''x'\|, \quad (1)$$

with $T^{-1} = (M^\top, -M^\top b)$ as the inverse transformation of $T = (M, b)$, $T'' = T^{-1}T'$ as another element of the symmetry group, and $\|.\|$ denoting the canonical Euclidean norm. We will make use of these properties when deriving a symmetrized Gaussian process model.

### B. Transformation of PES

Numerically, the total energy, as a function of the Cartesian coordinates of a rigid rotor molecule in an external structure, covers arbitrary orders of magnitude due to its numerous poles at geometries with overlapping atomic positions. Obviously, these regions are of little interest due to their non-physical energies. Typically, this is taken into consideration in Gaussian process models by using inverse coordinates as input for the regression model, since the positions and the energy output will then share a similar functional behavior.[50,51,56]

However, in order to make use of gradient information and the symmetry properties of the PES in the RR sub-manifold, we remain in a (constrained) Cartesian representation, parameterized by the RR coordinates, and apply a convenient transformation to the energy instead: For improved fitting performance and better resolution in PES regions of lower energy, we introduce a rational logarithmic transformation function of the form

$$\tau(E) = \frac{\log(1 + E_* - E_0)}{\log(1 + E_* - E_0) + (1/\varepsilon - 1)\log(1 + E - E_0)} \quad (2)$$

to switch from the actual energy to an abstract measure $\tau(E)$, where $E_0 \leq \min E \leq E_*$ and $0 < \varepsilon < 1$. The parameters $E_0$, $E_*$, and $\varepsilon$ can be chosen such that the region of interest is emphasized while the region $E > E_*$ is suppressed. The transformation function $\tau(E)$ is a monotonically decreasing function that maps $E_0$ to one and $E_*$ to $\varepsilon$. Figure 1 shows the effect of the rational logarithm transform for a two-dimensional cut through the PES of a He atom and a graphene model pore, the first benchmark system to be discussed in Sec. III. The energy surface depicted on the left is obtained when scanning over He positions within the plane of a rigid graphene model pore. The graphics in the center illustrates the transformation function with parameters $E_0 = \min E - 0.05$ eV, $E_* = E_0 + 0.5$ eV, and $\varepsilon = 0.1$. The transformed potential energy surface is shown on the right. Note how high energies are suppressed and low energies are emphasized. Moreover, the different energy scales are smoothed by the application of the logarithm.

Gaussian process regression is then applied to this transformed potential energy surface using a model that incorporates the symmetries of the rigid rotor sub-manifold. It proves useful to also add gradient information to the fitting strategy since many implementations of standard energy calculators do provide analytical gradients by default. In the following sections, we will derive a symmetry-enhanced Gaussian process regression model that is capable of incorporating space group symmetry operations as well as gradient information.

### C. Gaussian process regression

First, we briefly summarize the concept of Gaussian process regression and introduce a suitable notation. A Gaussian process is a collection of random variables, any finite number of which has a joint Gaussian distribution. This is represented by the notation $f(x) \sim \mathcal{GP}(m(x), k(x, x'))$, where $f$ is the function being modeled, $m(x)$ is the mean of the function, and $k(x, x')$ is the covariance function. The mean can be set to zero without loss of generality, which simplifies further discussion. When a finite set of evaluation points $X$ is considered, the function values $f$ at these points obey a joint Gaussian distribution of the form $f \sim \mathcal{N}(0, K)$, with $K_{ij} = k(x_i, x_j)$ as the covariance matrix. However, in most cases, one is not only interested in the prior distribution of the function but rather in exploiting the knowledge that the accessible training data provide for a guess of the underlying function. This is done by calculating the conditional distribution of the test data based on the training data. The training data are denoted as $(y, X)$, where $y = f + \epsilon$ is the noisy training data, and $\epsilon$ is independent, identically distributed Gaussian noise with mean 0 and variance $\sigma_n^2$. The test data are denoted by $(f_*, X_*)$. Defining $K_* = k(X_*, X_*)$ and $k_* = k(X, X_*)$, the joint distribution of training and test data is given by

$$\begin{bmatrix} y \\ f_* \end{bmatrix} \sim \mathcal{N}\left(0, \begin{bmatrix} K + \sigma_n^2 I & k_* \\ k_*^\top & K_* \end{bmatrix}\right). \tag{3}$$

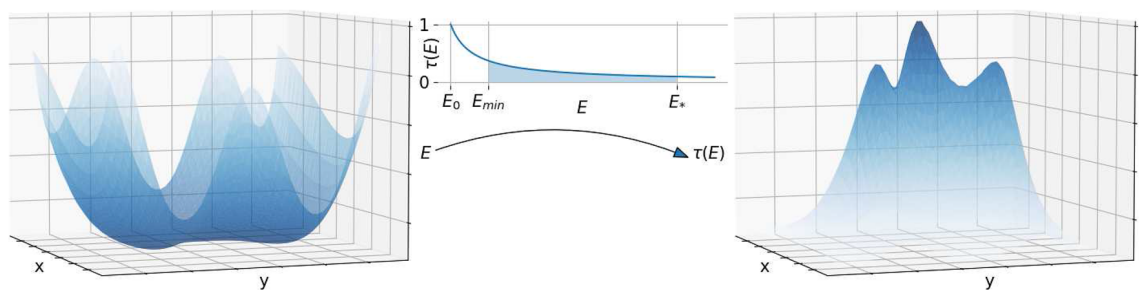

**FIG. 1.** Illustration of the rational logarithm transform for the PES of a helium atom in the pore plane of a graphene pore. In the left panel, the PES is shown. The right panel shows the transformed PES. The transformation function with parameters $E_0 = \min E - 0.05$ eV, $E_* = E_0 + 0.5$ eV, and $\varepsilon = 0.1$ is shown in between both plots, where the relevant energy range is shaded.

This yields the following conditional distribution of the test data:

$$(f_* | X_*, y, X) \sim \mathcal{N}\left(k_*^\top (K + \sigma_n^2 I)^{-1} y, K_* - k_*^\top (K + \sigma_n^2 I)^{-1} k_*\right), \tag{4}$$

which is referred to as the posterior or conditional distribution. Thus, we only have to decide on a prior distribution by choosing a suitable covariance function. Consequently, the following sections will be devoted to the derivation of a covariance function that takes symmetry as well as gradient information into account.

### 1. Inclusion of gradient information

Incorporating gradient information into Gaussian process regression can be achieved by exploiting that affine linear transforms of Gaussian processes are still Gaussian processes. This reasoning also applies to functions, which turns the latter finding into a much more general statement. For $f(x) \sim \mathcal{GP}(0, k(x, x'))$, the distribution of the first derivative is given by

$$\partial_x f(x) \sim \mathcal{GP}(0, \partial_x \partial_{x'} k(x, x')), \tag{5}$$

and the joint distribution of function values and gradients can be written as

$$\begin{bmatrix} f(x) \\ \partial_x f(x) \end{bmatrix} \sim \mathcal{GP}\left(0, \begin{bmatrix} k(x, x') & \partial_{x'} k(x, x') \\ \partial_x k(x, x') & \partial_x \partial_{x'} k(x, x') \end{bmatrix}\right), \tag{6}$$

as it is discussed in more detail in Refs. 37 and 38.

### 2. Inclusion of symmetry information

In a Gaussian process model, the choice of a prior distribution is crucial, and it is completely defined by the covariance function and the mean function. According to Mercer's theorem,[59] any positive definite covariance function can be related to a finite or infinite set of basis functions. Multiple basis functions ($\phi_i$) can be chosen for the function space, each of which results in a specific covariance function $k(x, x')$ by setting $f(x) = \sum_i w_i \phi_i(x)$, with multivariate, normally distributed weights $w \sim \mathcal{N}(0, \Sigma)$, i.e., with mean 0 and covariance matrix $\Sigma$, and calculating $k(x, x') = \mathbb{C}\text{ov}[f(x), f(x')]$.

A commonly used covariance function is the squared exponential kernel, also known as the Gaussian kernel, which is given by

$$k(x, x') = \sigma_f^2 \exp\left(-\frac{\|x - x'\|^2}{2\ell^2}\right), \tag{7}$$

with $\sigma_f^2$ relating to the data spread and $\ell$ denoting the correlation length. This covariance function corresponds to a Bayesian linear regression model with infinitely many Gaussian basis functions[59] and can easily be generalized to obey certain symmetry constraints as will be shown below. Note that, in our case, the distance featured in Eq. (7), $\|x - x'\|$, is the Euclidean distance of the molecular configurations $x, x'$ represented in the $3N_m$-dimensional Cartesian space. The same procedure can be applied to a symmetry-adapted Gaussian process model in $d$ dimensions, invariant under isometric affine linear symmetry transformations $T = (M, b)$ of a given space group. We represent the function $f(x)$ as a finite linear combination of $N^d$ symmetrized radial basis functions on an equidistant grid with $N$ gridpoints in each dimension

$$f(x) = \sum_{\alpha \leq N} w_\alpha \phi_\alpha(x), \tag{8}$$

with independent, identically distributed weights $w_\alpha \sim \mathcal{N}\left(0, \frac{\sigma_f^2}{N^d}\right)$ and a multiindex $\alpha$ labeling the position of the radial basis function $\phi_\alpha$. The symmetrized radial basis functions are of the form

$$\begin{aligned} \phi_\alpha(x) &= \frac{1}{(\sqrt{\pi}\ell)^d} \sum_i \exp\left(-\frac{\|x - T_i c_\alpha\|^2}{\ell^2}\right) \\ &= \frac{1}{(\sqrt{\pi}\ell)^d} \sum_i \exp\left(-\frac{\|T_i^{-1} x - c_\alpha\|^2}{\ell^2}\right), \end{aligned} \tag{9}$$

where a single Gaussian, centered at $c_\alpha$, is repeated at all positions through some operator $T_i$ of the group acting on $c_\alpha$, for all symmetries $i \in \{1, \ldots, N_{\text{sym}}\}$, with $N_{\text{sym}}$ denoting the total number of valid symmetry operations. Since the symmetry group is closed, invertible, and orthogonal, the symmetry operation $T_i$ can be moved to act on $x$ as shown in Eq. (1). Calculating the covariance and taking the limit to infinitely many gridpoints, i.e., $N \to \infty$, one obtains

$$\begin{aligned} k(x, x') &= \sigma_f^2 \sum_{ij} \exp\left(-\frac{\|T_i x - T_j x'\|^2}{2\ell^2}\right) \\ &= N_{\text{sym}} \sigma_f^2 \sum_m \exp\left(-\frac{\|x - T_m x'\|^2}{2\ell^2}\right), \end{aligned} \tag{10}$$

where one summation index has been removed by using Eq. (1) and noticing that each term appears $N_{\text{sym}}$ times in the sum. The multiplication by $N_{\text{sym}}$ can be absorbed into $\sigma_f^2$ by setting $\sigma_f^2 \to \sigma_f^2 / N_{\text{sym}}$. This way, the resulting kernel can be used for formally infinite symmetry groups as well, by taking the limit $N_{\text{sym}} \to \infty$. This may be useful when including lattice vectors of a periodically repeating external structure in the set of valid symmetry operations.

In the last step, we further introduce gradient information to the symmetrized kernel. For this purpose, we determine the first and second derivatives of the latter and rewrite the terms as single sums over symmetries. Defining

$$k_m(x, x') = k(x, T_m x') = \sigma_f^2 \exp\left(-\frac{\|x - T_m x'\|^2}{2\ell^2}\right), \quad (11)$$

by neglecting the symmetry factor $N_{\text{sym}}$, we get

$$\begin{aligned}
k(x, x') &= \sum_m k_m(x, x') \\
\partial_x k(x, x') &= -\frac{1}{\ell^2} \sum_m k_m(x, x')(x - T_m x') \\
\partial_{x'} k(x, x') &= -\frac{1}{\ell^2} \sum_m k_m(x, x')(x' - T_m^{-1} x) \\
\partial_x \partial_{x'} k(x, x') &= \frac{1}{\ell^4} \sum_m k_m(x, x')\bigl(\ell^2 M_m + (x - T_m x') \\
&\quad \otimes (x' - T_m^{-1} x)\bigr),
\end{aligned} \quad (12)$$

with $\otimes$ denoting the outer product of two vectors. This kernel incorporates symmetry and gradient information. We note that the exploitation of symmetry information for PES fitting was already proposed by Bartók et al.,[55] and a similar approach was presented by Chmiela et al. for the exploitation of symmetry in combination with kernel ridge regression.[53,54]

### 3. Hyperparameter optimization

The choice of the length-scale parameter in the covariance function can have a significant impact on the accuracy of any Gaussian process regression model, especially in cases where gradient information is used. Therefore, an optimal choice of the length scale is crucial for ensuring reliable PES emulation. Its optimization, along with all other hyper-parameters of the model, can be achieved in various ways. A common approach is to optimize the log-likelihood of the model. For a squared exponential kernel and (noisy) training data $(y, X)$, the log-likelihood is given by

$$\begin{aligned}
\log p(y|X) = &-\frac{1}{2} y^\top \bigl(K(X, X) + \sigma_n^2 I\bigr)^{-1} y \\
&- \log\bigl(\det\bigl(K(X, X) + \sigma_n^2 I\bigr)\bigr) - \frac{n}{2} \log(2\pi).
\end{aligned} \quad (13)$$

This approach, however, leads to unstable results in our case because the obtained hyperparameters are not "optimal" in terms of the prediction error. Benchmark calculations indicate that training data points are well described, but the prediction of test data points remains insufficient. We assume that this behavior is due to different length-scales appearing on the PES. However, including a variable length-scale as a direct remedy would introduce a very high level of complexity to the model.

To overcome this issue, a cross-validation approach can be employed instead. Since the goal is to obtain a good fit on the potential energy surface, a direct measure of the fit fidelity, such as the mean squared prediction error (MSPE) of the training data, can be used to optimize the hyperparameters. For a set of training data points $X = (x_i)_{i=1,\ldots,N}$ and $y = (y_i)_{i=1,\ldots,N}$, the MSPE is obtained via

$$\rho_{\text{MSPE}} = \sum_{i=1}^{N} (y_i - y_{\neg i})^2, \quad (14)$$

with $y_{\neg i}$ denoting the predicted values using all the $i$th data points. The optimal hyperparameters can be obtained by minimizing the MSPE, e.g., via gradient descent. This approach is more stable than the maximization of the log-likelihood, as it directly measures the fidelity of the fit given the training data available. In addition, it is consistent with the specification of iterative predictive methods that are designed to be more robust.[59,60] By minimizing the MSPE, we can ensure that the length-scale parameter is chosen to be optimal for a given PES in terms of energy prediction.

### 4. Active learning

Active learning aims for an optimized choice of data points to be added to an existing set based on current knowledge. This feature can be easily added to GPR. Typically, one chooses a point of maximum information gain for the model as the location for the additional training point evaluation. This can be achieved by minimizing an acquisition function that quantifies the uncertainty of the model. Commonly used acquisition functions range from simple variance estimators[57] to expected information gains.[58] A trade-off between exploration and exploitation is then aspired, i.e., by choosing points where the model is uncertain vs choosing points that are likely to improve the model. For the task at hand, this is achieved by the acquisition function

$$\mu(x) = -\mathbb{V}\text{ar}[f(x)|y, X]\bigl(\epsilon + \mathbb{E}[f(x)|y, X]^2\bigr), \quad (15)$$

where variance and expectation are taken from the GPR fit function with training data $(y, X)$ and $0 < \epsilon \approx 10^{-3}$ in our case to avoid that $\mu(x)$ vanishes if $f(x)$ is zero. The minimization of this function will deliver a point that exhibits high variance and a high function value, which results in a point search preferably sampling high value regions. The incorporation of gradient information is highly beneficial for the active learning process. Information on the slope of the given PES reveals more details of the underlying function, resulting in a more efficient search in high-value regions.

### D. Technical details of the GPR implementation

The symmetry groups of the molecule and its external structure are obtained via the symmetry analyzer implemented in the pymatgen module.[61,62] Symmetry elements are translated into their $3N_m$-dimensional representations acting on the Cartesian coordinates of the molecule. The proper symmetry operations for mapping the rigid rotor sub-manifold into itself are selected by calculating the determinant of the respective transformation matrix. In addition, several useful routines of the atomic simulation environment (ASE),[63] a convenient Python module, are used.

An important aspect of our regression scheme is that we remain in a Cartesian description of the RR sub-manifold, since the inclusion of symmetry operations and gradient information is most convenient in this representation. RR coordinates are only used for the sake of a symmetry-adapted parameterization, i.e., to obtain training points within the RR sub-manifold. Therefore, the choice of RR coordinates is of minor importance for our method, as the regression is done in Cartesian space. This extends to the active learning scheme, which is not influenced by any ambiguities or discontinuities in the RR coordinates.

The symmetry-adapted GPR is initialized with a custom number of random points. Additional training points are then selected according to the active learning strategy outlined above or are chosen randomly. The minimization of the respective search function is performed using the constrained basin hopping algorithm as it is implemented in the scipy.optimize library, working as a global minimizer in the region of interest.[64] After each $n$ iterations, the hyperparameters are optimized via the minimization of the mean squared prediction error. Note that parameters $E_0$ and $E_*$ of the transformation function are updated during fitting in order to increase the performance and flexibility of the transformation. This way, the relevant energy region does not have to be known beforehand but is found automatically in the course of the regression process. This also means that the transformed surface $\tau(E)$ will change its shape during the regression. Note, however, that the latter does not influence the shape of the fitted PES. An overview of Algorithm 1 in pseudocode is given below.

For all PES evaluations, we use GFN-FF as our external energy calculator,[65] a generic force field recently developed by the Grimme group with the aim to enable fast structure optimizations and molecular-dynamics simulations of large molecular systems. The GFN-FF force field includes asymptotically correct bond stretch, bond angle, and torsional terms for covalent interactions, three-body corrections, a Gaussian-type potential to mimic bond breaking where necessary, diffusion corrections, and a topology-based charge model that introduces partial polarizability. It allows a full evaluation of PES surfaces due to its computational minimalism, but can be considered as a highly suitable low-cost approximation of typical PES surfaces evaluated at a much higher level of theory in terms of relative energy predictions and actual surface complexity.

The minimal computational effort of single point evaluations with this method allows for high precision in all integral-based estimates of the PES fit quality, which typically involve thousands of energy and gradient evaluations. We note that this is the only reason for the employment of a force-field ansatz in this manuscript. As shown in Sec. III, our method is capable of delivering accurate PES fits based on less than 100 data points if symmetries and active learning can be exploited, which enables the usage of much more expensive calculators such as coupled cluster methods or wavefunction-based multi-reference techniques e.g., commonly applied to electronically excited states.

## III. RESULTS AND DISCUSSION

We test our symmetry- and gradient-enhanced GPR algorithm on four different molecular scenarios, which have been chosen in order to investigate the impact of exploitable symmetries in a systematic way. In all cases, we compare standard GPR to all possible extensions by adding symmetry information (sym.) and gradient information (grad.) with and without active learning (al.). The proposed method, of course, inherits all three methodological extensions (GPR + sym. + grad. + al.).

The first three cases are examples of molecular sieving via porous graphene, simulated via a model pore obtained by the removal of carbon atoms from a single layer of graphene, followed by the replacement of the unsaturated carbon atoms by nitrogen. Chosen not only for a controlled benchmarking of our method, this type of substitution in porous carbon structures has been investigated in the past,[9] in particular, with respect to electrostatic effects in the context of Lewis-type acid–base interactions between pyridine-like functionalizations and carbon dioxide.[66,67] Nitrogen-containing heterocyclic aromatic molecules are also interesting linker molecules in zeolitic imidazole frameworks (ZIFs), a type of metal–organic framework with zeolitic topology and possible applications for $CO_2$ sequestration.[68–73] The first situation concerns a single helium atom and an N-functionalized model pore obtained by the removal of five carbon atoms. This system has only four symmetries in total. The second, related test scenario involves molecular hydrogen and an N-saturated graphene pore of approximately rectangular shape with edges of about 2 Å length. A third membrane-inspired problem, this time with significantly higher symmetry, concerns methane and a six-fold rotationally symmetric graphene pore, obtained by the removal of a whole benzene ring, again followed by nitrogen substitution. The last testing case, slightly different in nature, is concerned with the intermolecular interaction between $CH_4$ and $N_2$, and has been selected to demonstrate the principal applicability of our code and for the problems related to solubility, mesophase systems, and molecular clustering.

The accuracy of our algorithm is evaluated by the comparison of the GPR fit to the actual PES. Since the GFN-FF reference PES can be evaluated with little effort at any point,[65] the quality of the GPR method can be determined *a posteriori* simply by requesting energy evaluations at points of the PES that are yet unknown to the algorithm. The actual PES energy value of a certain geometry and its fitted value are denoted as $E$ and $\hat{E}$, respectively.

Since the deviations at high energy values are less relevant for most applications, it makes sense to opt for an alternative, weighted scalar quantity instead. Therefore, two new measures of energy deviation are introduced as follows: The first is the weighted infinity norm

$$\|\Delta E\|_\infty = \sup_{x \in \Omega} \left\{ |\hat{E} - E| e^{-\beta(E - E_{\min})} \right\}, \quad (16)$$

where the normalized Boltzmann weight $e^{-\beta(E - E_{\min})}$, with $E_{\min}$ as the minimal energy, provides a natural choice for a weighted point

---

**ALGORITHM 1.** Symmetrized Gaussian process regression for PES fitting.

**Input:** external structure, molecule, *tol*
**Output:** emulated PES
  construct the symmetric GPR model $\hat{E}$ with valid symmetry operations of the RR sub-manifold
  $X_0 \leftarrow$ set initial training data
  $E_0 \leftarrow \min E(X_0) - 0.05$ eV
  $\varepsilon \leftarrow 0.1$
  $Y_0 \leftarrow \tau(E(X_0))$
  **while** not converged **do**
    $x_+ \leftarrow \arg\min \mu(x)$ or random
    add $((x_+, \tau(E(x_+)))$ to the training data
    $E_0 \leftarrow \min_{\text{training data}} E - 0.05$ eV
    $E_* \leftarrow E_0 + 0.5$ eV
    lengthscale $\leftarrow \arg\min \rho_{\text{MSPE}}$
  **end while**

distance measure. Another natural measure of "distance" in energy is given by the weighted averaged quadratic distance

$$\|\Delta E\|_2 = \sqrt{\int_\Omega (\hat{E} - E)^2 p(E) dx}, \quad (17)$$

where $p(E) = \frac{e^{-\beta E}}{Z}$ is the Boltzmann probability at inverse temperature $\beta$ with partition sum $Z = \int_\Omega e^{-\beta E} dx$. Essentially, this corresponds to the thermal expectation value of the quadratic energy deviation.

Note that these measures are not accessible "on-the-fly" during active learning due to their *a posteriori* character. Both rely on a large number of randomly chosen point evaluations on the GFN-FF reference PES since the averaged quadratic distance is determined via Monte Carlo integration. However, both distance measures can be used to estimate the fit quality by comparison to previous iterations and may be used in future implementations to define convergence criteria for an automated PES fitting. For this purpose, the maximum norm will be preferred as it does not require a potentially time-consuming multi-dimensional integration of the partition sum.

### A. Helium in the graphene pore

The first testing case, a problem, e.g., relevant for membrane-based isotopic separation,[8,74] bears a minimal chance for the exploitation of symmetry by the GPR algorithm but offers meaningful insights via the possibility of a graphical PES depiction. The molecular system, consisting of a single He atom and a N-functionalized, graphene-derived model pore, is shown in Fig. 2. We constrain the investigation to in-plane positions of the He atom, which reduces the total number of degrees of freedom to two in the case of a frozen pore geometry. An evaluation of this area is particularly interesting in the context of extended transition state methods,[25,75,76] which go beyond single trajectory-based treatments such as Eyring theory. The corresponding PES has already been presented in Fig. 1 to show the effect of the energy transform function $\tau(E)$. In addition, the obvious mirror symmetry, which is of no help for single point evaluations exactly within the pore plane, the structure obtained by the removal of five carbon atoms features a $C_2$ axis. This is the only symmetry element that can be exploited by the GPR algorithm in this constrained, two-dimensional evaluation.

A direct comparison of convergence for the various GPR methods on this system is shown in Fig. 3, which shows a double-logarithmic graph of the energy deviations between the fitted and the reference PES, calculated in the two ways discussed above, as a function of the number of PES single point evaluations. For the sake of a direct comparison, the same type of graph will be presented for all four benchmarking scenarios. In this first testing case, a significant improvement is observed as soon as gradient information is included, while the inclusion of symmetry offers only a minor advantage. The inclusion of active learning improves the performance significantly in the symmetrized and unsymmetrized regressions. The unsymmetrized regression benefits even more from the addition of active learning by more than an order of magnitude. Since effectively only one symmetry can be exploited, the GPR model with gradient information and active learning outperforms the symmetrized versions without active learning. However, note that the addition of symmetry information to an already gradient-enhanced GPR variant with active learning still improves the accuracy by a factor of 5, and spectroscopic accuracy, which we define as deviations below 1 meV (8 cm$^{-1}$ or $4 \times 10^{-5}$ hartree), is reached already after about 50 single point evaluations.

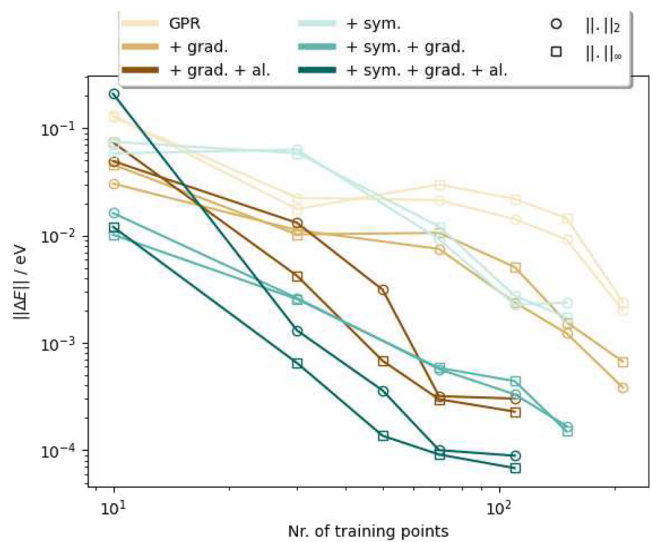

**FIG. 3.** Convergence behavior of the six GPR variants for benchmark scenario 1, He in porous graphene, using the mean quadratic distance (circles) and the maximal weighted distance (squares) as a measure of quality.

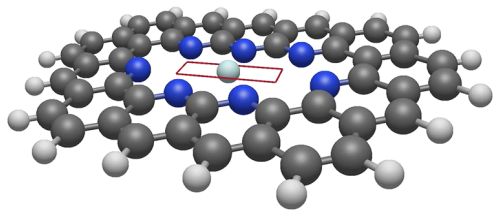

**FIG. 2.** A nitrogen-functionalized graphene model pore of $C_{2h}$ symmetry with a single helium atom located within the pore plane. The two-dimensional PES scanning area is indicated by red lines.

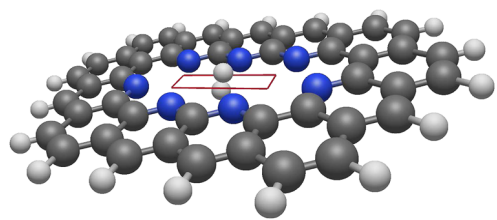

**FIG. 4.** A H$_2$ molecule residing in the plane of an N-functionalized graphene nanopore with $C_2$ symmetry. The PES scanning area is indicated by red lines, a two-dimensional rectangular volume for the center-of-mass coordinates of the hydrogen molecule. Together with two angular coordinates describing its respective orientation, a four-dimensional PES needs to be fitted in this scenario.

## B. H$_2$ in a graphene pore

In the second testing scenario, we remain with the same pore model but replace the He atom with molecular hydrogen, which introduces two additional degrees of freedom describing the orientation of the H$_2$ internuclear axis with respect to the pore. For the sake of a convenient graphical representation, we restrain ourselves again to H$_2$ center-of-mass positions within the pore plane, as illustrated in Fig. 4. In total, this leaves us with a four-dimensional PES spanned by the $x$ and $y$ center-of-mass positions and two Euler angles. An overview of this already rather complicated PES is provided in Fig. 5, where each of the nine subplots represents a PES cut (energy as a function of the $x$ and $y$ positions) for a certain orientation of the H$_2$ molecule, indicated in the upper right corner. Red surfaces show the GPR-based approximation, including gradient and symmetry information, while blue surfaces correspond to the reference PES of the system. It is remarkable that a useful approximation in all four dimensions can already be achieved with only 52 single point evaluations. The reason for this efficiency lies in just one additional symmetry feature that can be exploited: the C$_2$ rotation around an axis perpendicular to the H$_2$ internuclear axis.

In Fig. 6, we again compare the convergence behavior of the four GPR variants. It clearly demonstrates the efficiency of the combined algorithm, which is the only method reaching deviations below $10^{-2}$ eV after about 100 single point evaluations. It is also visible that gradient information alone can be problematic at the start, i.e., in cases of minimal information, while the addition of symmetry information tends to stabilize the convergence behavior and leads to an even better performance eventually. For this scenario, where more symmetries can be exploited, even the symmetrized GPR without gradient information performs equally well as the unsymmetrized GPR with gradient information and active learning. However, the inclusion of active learning again improves performance significantly.

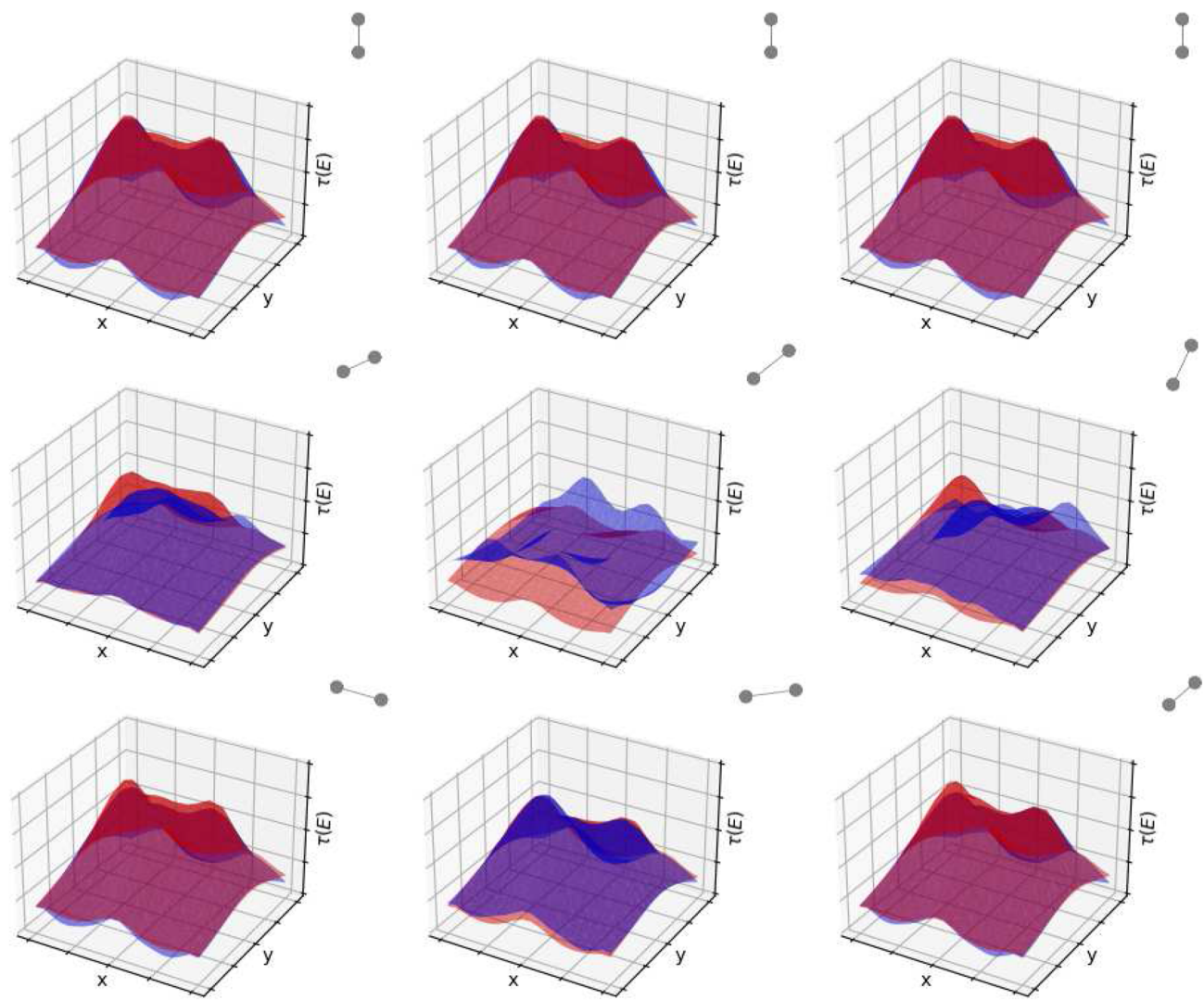

**FIG. 5.** Cuts through the PES of test case 2, H$_2$, in the vicinity of porous graphene. Predicted energies are shown in red, and the actual PES is printed in blue. The H$_2$ is kept in the pore plane; energies are plotted as a function of the $x$ and $y$ coordinates of the molecular center of mass. Each of the nine plots corresponds to a different orientation of H$_2$ with respect to the pore, indicated by a small inset in each upper right corner. Only 52 training points (in total) were used to generate the PES cuts.

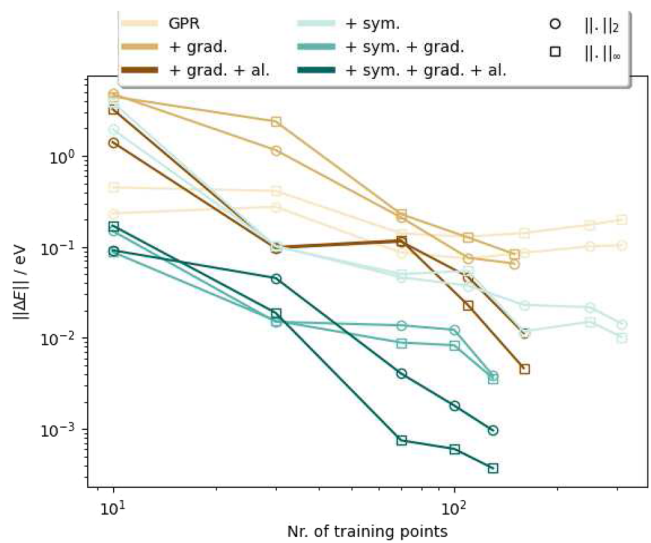

**FIG. 6.** Convergence behavior of the six GPR variants for benchmark scenario 2, $H_2$ in porous graphene, using the mean quadratic distance (circles) and the maximal weighted distance (squares) as a measure of quality.

### C. $CH_4$ in box around graphene pore

As a next step, we increase the dimensionality of the problem as well as the number of available symmetry elements. We investigate the case of a methane molecule in the vicinity of a graphene pore obtained by the removal of six C atoms, again followed by the substitution of unsaturated carbon atoms by nitrogen. This pore is a representative of the $C_{6h}$ molecular point group, while the highly symmetric $CH_4$ belongs to $T_d$. The former introduces 12 symmetry elements, the latter 24. Contrary to the previous two tests, we will be evaluating the PES in all six dimensions of the RR approximation. The three-dimensional box of allowed center-of-mass positions of the $CH_4$ molecule is indicated in Fig. 7 by red lines.

The convergence behavior for this molecular scenario is shown in Fig. 8. The first insight is the total failure of gradient-enhanced GPR, which is unable to offer even the slightest improvement within the first 100 single point evaluations. It performs even worse than standard GPR, a phenomenon we attribute to the high number of symmetry elements leading to a similarly large number of equivalent local minima with a "distracting" effect on the gradient-driven GPR.[77] The purely symmetry-enhanced GPR, on the other hand, is able to reduce the deviations by almost two orders of magnitude—a clear consequence of the strongly reduced effective volume in configuration space. The convergence toward useful energy predictions is slower than in previous scenarios due to the high dimensionality. However, the combination of symmetry- and gradient-information yields predictive improvements by another order of magnitude.

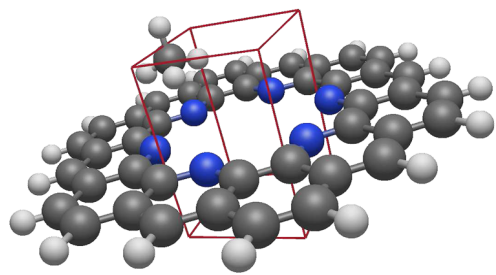

**FIG. 7.** A $CH_4$ molecule in the vicinity of a hexagonal graphene nanopore. The PES is evaluated for center-of-mass positions of methane that lie inside the red box. Together with the three Euler angles, the PES is evaluated in all six dimensions of the RR approximation.

Note that, even with just 30 training points in total, a meaningful fit with deviations in the order of a few meV can be achieved for this complicated six-dimensional PES. In addition, for this benchmark scenario, the implementation of the active learning scheme, coupled with the utilization of symmetries, gives rise to a substantial enhancement of method efficiency. An increase in accuracy by an order of magnitude becomes evident in comparison to uninformed point sampling, as the knowledge of minimal energy configurations is essential for gradient-enhanced GPR.

Additional insights can be gained from an analysis of the actual PES changes that occur in the course of an ongoing GPR evaluation. Despite the six-dimensional character, the ability of the symmetry- and gradient-enhanced GPR methods to learn actual surface features can be studied through convenient cuts: Fig. 9 shows the improvement via contour plots of a two-dimensional sub-manifold of the PES (lower row), obtained via scanning over two Euler angles, while all other remaining coordinates, i.e., the center-of-mass position of $CH_4$ and the third angle, are kept fixed. Note that transformed energies are dimensionless. For the sake of completeness, the corresponding cut through the transformed reference PES, $\tau(E)$, is also given for each intermediate result (upper row), since the reference PES also depends on $n$ via the transformation parameters [see Eq. (2)], which might change slightly in the active learning process. The incorporation of symmetry allows the GPR procedure to catch even the fine details of this complicated functional dependence with about 30 training points.

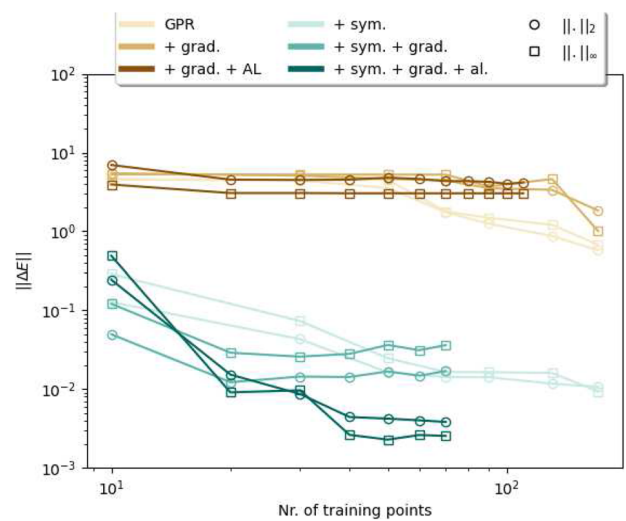

**FIG. 8.** Convergence behavior of the six GPR variants for benchmark scenario 3, $CH_4$ in the vicinity of a hexagonal pore, using the mean quadratic distance (circles) and the maximal weighted distance (squares) as a measure of quality.

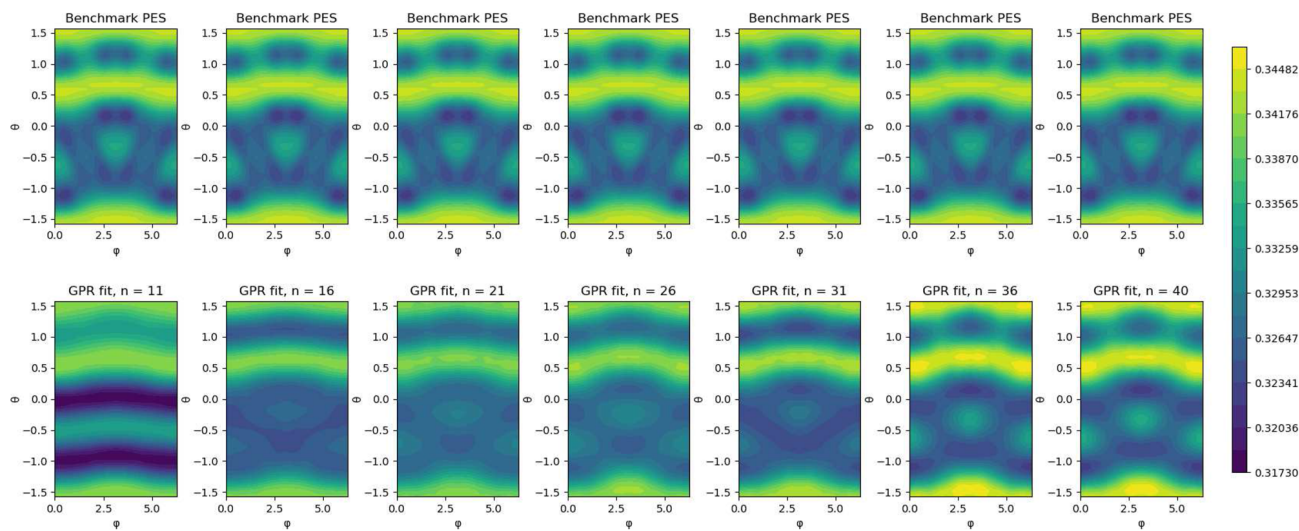

**FIG. 9.** Illustration of the convergence process for test case 3, the PES of methane in the vicinity of porous graphene. Two-dimensional cuts through the PES are compared at various stages of the algorithm, with $n$ denoting the number of training points. The two degrees of freedom correspond to rotations with respect to angles $\theta$ and $\phi$. The lower row shows the prediction (GPR + symm. + grad.), the upper the actual PES.

### D. Intermolecular interaction

Finally, we test our GPR model on a slightly different situation, investigating the inter-molecular interaction of $CH_4$ and $N_2$,

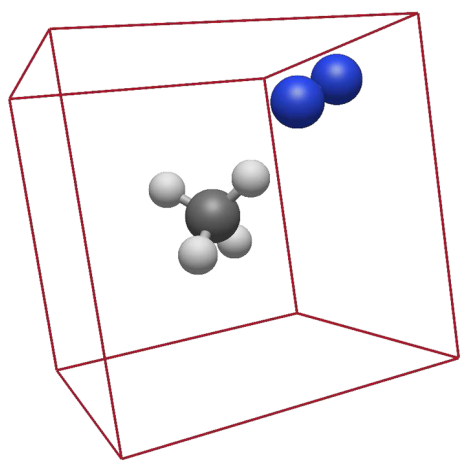

**FIG. 10.** $N_2$ molecule in the vicinity of a $CH_4$ molecule located at the center of the configuration space indicated by red lines.

representatives of the $T_d$ and $D_{\infty h}$ molecular point groups, respectively. The scenario is shown in Fig. 10, again with the simulation box indicated by red lines. The methane molecule is kept fixed at the center, while the position and orientation of the nitrogen molecule are varied in the remaining five degrees of freedom. The corresponding convergence behavior of the four GPR variants is shown in Fig. 11. Similarly to the previous scenario, the convergence is slightly slower for all methods due to the higher dimensionality. Again, the symmetry- and gradient-enhanced GPR method with active learning performs best, reaching an accuracy well below 10 meV after 100 single point evaluations. Similar to the other testing cases, the inclusion of symmetry information has a stabilizing effect on the overall convergence behavior, while the usage of gradient information alone leads to much less accuracy gain. In the case of symmetrized GPR, active learning immediately enhances the performance, whereas the unsymmetrized GPR does not benefit initially from active learning but necessitates a larger number of training points. This is explained by the fact that the active learning scheme favors minima in the local surroundings of already existing training points; hence, if the sampling volume is large, a sufficient exploration of the PES necessitates more single point evaluations due to this intrinsic preference.

Note that this last testing case comprises a much larger volume in real space than the previous scenarios, i.e., a much larger box for the placement of the "mobile" $N_2$ relative to the static $CH_4$ molecule. The success of enhanced GPR also in this case suggests future tests on even larger molecular environments, the implementation of method mixing to account for short- and long-range interactions at different levels of theory, and the extension or generalization toward periodic systems. For the case of $CH_4$ and $N_2$, a comparison can be made between our results and those obtained by Uteva *et al.*, who have examined this specific intermolecular interaction scenario.[50] With about 100 training points, our accuracy lies in the meV range, which is comparable in its order of magnitude.

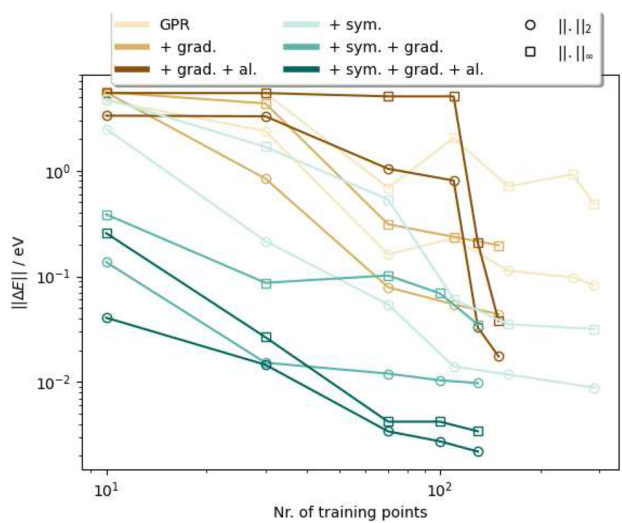

**FIG. 11.** Convergence behavior of the six GPR variants for benchmark scenario 4, the inter-molecular interaction of $CH_4$ and $N_2$, using the mean quadratic distance (circles) and the maximal weighted distance (squares) as a measure of quality.

## IV. CONCLUSION

In this study, we tested the performance of a symmetry- and gradient-enhanced GPR model for PES fitting of molecular systems where single molecules are weakly interacting with a stationary molecular environment. In contrast to more generic implementations for inter-molecular[50,51] or intra-molecular[53,54] interactions, our method intends to address molecular scenarios characterized by rigid, freely moving molecular objects in a periodic, symmetric environment. Four systems, including three applications of molecular sieving via nitrogen-functionalized porous graphene and the inter-molecular interaction of $CH_4$ and $N_2$, have been chosen as benchmark scenarios of varying symmetry and dimension. In all cases, a substantial improvement standard GPR can be observed by taking the following measures in algorithm design:

- use a rational, logarithmic transformation of the PES that emphasizes interesting, low energy regions and mitigates divergent behavior near poles
- include gradient information for better extrapolation of training data, leading to enhanced performance of the active learning procedure
- include symmetry information via a symmetry-adapted Gaussian kernel to reduce the effective volume of configuration space
- employ an active learning strategy aiming for an optimal exploration and exploitation of the configuration space by favoring minimal energy values and high variance

Our study demonstrates the effectiveness of using symmetry- and gradient-enhanced GPR models with active learning for PES fitting. Large improvements in convergence behavior and accuracy can be observed, often providing PES fits of spectroscopic accuracy, i.e., deviations below 1 meV, with less than 100 single point evaluations. This enables the use of high-level energy predictors such as density functional theory or even coupled cluster or multi-reference methods to generate training data, which opens the possibility to still apply costly electronic structure theory even to larger molecular systems. The proposed method is particularly well-suited for PES approximations of confined molecules in external structures with complicated regions of interest and smaller volumes, i.e., for scenarios of molecular adsorption, separation, sequestration, or storage on surfaces, membranes, or porous materials such as metal organic frameworks and zeolites. A first, successful test on intermolecular interactions also suggests future applications to problems of solubility, mesophase systems, and studies on molecular clusters.

## ACKNOWLEDGMENTS

Financial support by the Austrian Science Fund (FWF) under Grant No. P 29893-N36 is gratefully acknowledged. We further thank the IT services (ZID) of the Graz University of Technology for providing high performance computing resources, and technical support.

## AUTHOR DECLARATIONS

### Conflict of Interest

The authors have no conflicts to disclose.

### Author Contributions

**Johannes K. Krondorfer**: Conceptualization (equal); Formal analysis (lead); Investigation (equal); Methodology (equal); Software (equal); Validation (equal); Writing – original draft (equal); Writing – review & editing (equal). **Christian W. Binder**: Conceptualization (equal); Formal analysis (equal); Funding acquisition (lead); Investigation (equal); Methodology (equal); Project administration (lead); Software (equal); Supervision (lead); Validation (equal); Writing – original draft (equal); Writing – review & editing (equal). **Andreas W. Hauser**: Conceptualization (equal); Formal analysis (equal); Funding acquisition (lead); Investigation (supporting); Methodology (supporting); Project administration (lead); Supervision (lead); Writing – original draft (equal); Writing – review & editing (equal).

## DATA AVAILABILITY

The data that support the findings of this study are available from the corresponding author upon reasonable request.